%% file: main_arxiv.tex
\documentclass[a4paper,UKenglish,cleveref, autoref]{article}
\usepackage[algoruled,vlined,linesnumbered,algo2e,lined,noend]{algorithm2e}
\usepackage{graphicx}
\usepackage{makeidx}
\usepackage{amsmath, amssymb}
\usepackage{bm}
\usepackage{url}
\usepackage{wrapfig}
\usepackage{multirow}
\usepackage{authblk}
\usepackage{hyperref}
\usepackage{cleveref}
\usepackage{geometry}
\usepackage{color}
\newtheorem{theorem}{Theorem}
\newtheorem{definition}{Definition}

\newcommand{\occ}{\mathit{occ}}

\newcommand{\ksc}{\mathit{KSC}}
\newcommand{\kc}{\mathit{KC}}
\newcommand{\esp}{\mathit{ESP}}
\newcommand{\easp}{\mathit{EASP}}

\makeatletter
\newcommand{\figcaption}[1]{\def\@captype{figure}\caption{#1}}
\newcommand{\tblcaption}[1]{\def\@captype{table}\caption{#1}}
\makeatother

\title{Detecting $k$-(Sub-)Cadences and \\ Equidistant Subsequence Occurrences}

\author[1]{Mitsuru~Funakoshi}
\author[1]{Yuto~Nakashima}
\author[1]{Shunsuke~Inenaga}
\author[1]{Hideo~Bannai}
\author[1]{Masayuki~Takeda}
\author[2]{Ayumi~Shinohara}
\affil[1]{Department of Informatics, Kyushu University, Japan.

\{mitsuru.funakoshi, yuto.nakashima, inenaga, bannai, takeda\}@inf.kyushu-u.ac.jp}
\affil[2]{Graduate School of Information Sciences, Tohoku University, Japan.

ayumis@tohoku.ac.jp}
\date{}
\begin{document}

\maketitle

\begin{abstract}
  \fontsize{9}{12}\selectfont
  The equidistant subsequence pattern matching problem is considered. Given a pattern string $P$ and a text string $T$, we say that $P$ is an \emph{equidistant subsequence} of $T$ if $P$ is a subsequence of the text such that
  consecutive symbols of $P$ in the occurrence are equally spaced.
  We can consider the problem of
  equidistant subsequences as generalizations of (sub-)cadences.
  We give bit-parallel algorithms that yield $o(n^2)$ time algorithms for finding
  $k$-(sub-)cadences and equidistant subsequences.
  Furthermore, $O(n\log^2 n)$ and $O(n\log n)$ time algorithms,
  respectively
  for equidistant and Abelian equidistant matching
  for the case $|P| = 3$, are shown.
  The algorithms make use of a technique that was recently introduced which can efficiently compute convolutions with linear constraints.
\end{abstract}

\input{intro_arxiv}
\input{preliminaries_arxiv}
\input{k-sub-cadences_arxiv}
\input{k-cadences_arxiv}
\input{equidistant_subsequence_pattern_matching_arxiv}
\input{equidistant_subsequence_pattern_of_length_three_arxiv}

\clearpage
\bibliography{ref}

\end{document}

%% file: intro_arxiv.tex
\section{Introduction}

Pattern matching on strings is a very important topic in string
processing.
Usually, strings are regarded and stored as one dimensional sequences
and many pattern matching algorithms have been proposed
to efficiently find particular substrings occurring in them~\cite{KnuthMP77,BoyerM77,DBLP:conf/stringology/FaroLBMM16,Horspool1980,GALIL1983280,CrochemorePerrin91}.
However, when one is to view the string/text data on paper or on a screen,
it is usually shown in two dimensions:
the single dimensional sequence is displayed in several lines folded by
some length.
It is known that the two dimensional arrangement can be used to embed hidden messages,
and/or cause occurrences of unexpected or unintentional messages in the text.
A common form for such an embedding is to consider the occurrence of a
pattern in a linear layout: vertically or possibly diagonally along
the two dimensional display.

For example, there was a (rather controversial)
paper~\cite{witztum94:_equid_letter_sequen_book_genes} on the so
called Bible Code, claiming that the Bible contains statistically
significant occurrences of various related words, occurring vertically
and/or diagonally, in close proximity.
Furthermore, there was an incident with a veto letter by
the California State Governor~\cite{matier09:_did_schwar};
Although it was considered a ``weird coincidence'',
the first character on each line of the letter could be connected
and interpreted as a very provocative message.
In Japanese internet forums,
there was a culture of actively using these techniques, referred to as
``tate-yomi''(vertical reading) and
``naname-yomi'' (diagonal reading),
where the author of a message purposely embeds a hidden message in
his/her post.
Most commonly, the author will write a message that praises some
object or opinion in question, but embed a message with a completely
opposite meaning bearing the author's true intention.
The hidden message can be recovered by reading the text message
vertically or diagonally from some position, and is used as form of
sarcasm, as well as a clever method to mock those who were unable to
\emph{get} it.

Assuming that the text is folded into lines of equal length,
vertical or diagonal occurrences of the
pattern in two dimensions can be regarded as a subsequence of the
original text, where the distance between each character is equal.
We call the problem of detecting such occurrences of the pattern as the \emph{equidistant subsequence matching} problem.
To the best of the authors'
knowledge,
there exist only publications concerning
the statistical properties of the occurrence of equidistant subsequence
patterns, mainly with the so called Bible Code.

Recently, a notion of regularities in strings called
\emph{(Sub)-Cadences},
defined by equidistant occurrences of the same character,
was considered by Amir et al.~\cite{Cadences_Amir}.
A $k$-sub-cadence of a string can be viewed as an occurrence of an
equidistant subsequence of length $k$ that consists of the same character.
A $k$-sub-cadence is a $k$-cadence,
if the starting position is less than or equal to $d$ and
the ending position is greater than $n-d$, where $d$ is the
distance between each consecutive character occurrence
and $n$ is the length of the string.
To date, algorithms for detecting
anchored cadences (cadences whose starting position is equal to $d$),
$3$-(sub-)cadences,
and $(\pi_1, \pi_2, \pi_3)$-partial-$3$-cadences
(an occurrence of an equidistant subsequence that can become a cadence by changing at most all but 3 characters)
have been proposed~\cite{Cadences_Amir, Funakoshi_STACS2020}.
However, no efficient algorithm for detecting
$k$-(sub)-cadences for arbitrary $k~(1\leq k \leq n)$
is known so far.

In this paper, we present counting algorithms for $k$-sub-cadences, $k$-cadences, equidistant subsequence patterns of length $m$ and length $3$,
and equidistant Abelian subsequence patterns of length $3$.
Table~\ref{tab:complexity} shows a summary of the results.
All algorithms run in $O(n)$ space.
Furthermore, we present locating algorithms for $k$-sub-cadences, $k$-cadences, and equidistant subsequence patterns of length $m$.
The time complexities of these algorithms can be obtained by adding $\occ$ to the second term inside the minimum function of each time complexity of the counting algorithm.
To the best of the authors'
knowledge, these are the first $o(n^2)$ time algorithm for $k$-(sub)-cadences and equidistant subsequence patterns.
Unless otherwise noted, we assume a word RAM model with word size
$\Theta(\log n)$,
and strings over a general ordered alphabet.

\begin{table}[htb]
  \begin{center}
    \begin{tabular}{|c||c|c|} \hline
      Counting time & For a constant size alphabet & For a general ordered alphabet \\ \hline \hline
      $k$-sub-cadences & $O\left(\min\left\{\frac{n^2}{k},\frac{n^2}{\log n}\right\}\right)$ & $O\left(\min\left\{\frac{n^2}{k},\frac{n^2 \sqrt{k}}{\sqrt{\log n}}\right\}\right)$ \\ \hline
     $k$-cadences & $O\left(\min\left\{\frac{n^2}{k},\frac{1}{\log n}\left(\frac{n^2}{k^2}+kn\right)\right\}\right)$& $O\left(\min\left\{\frac{n^2}{k},\frac{n \sqrt{k}}{\sqrt{\log n}}\sqrt{\frac{n^2}{k^2}+kn}\right\}\right)$ \\ \hline
    \end{tabular}

    \vspace{5mm}

    \scalebox{0.98}{
    \begin{tabular}{|c||c|} \hline
      Counting time & For a general ordered alphabet \\ \hline \hline
      Equidistant subsequence pattern & $O\left(\min\left\{\frac{n^2}{m},\frac{n^2}{\log n}\right\}\right)$ \\ \hline
     Equidistant subsequence pattern of length three & $O(n \log^2 n)$ \\ \hline
     Equidistant Abelian subsequence pattern of length three & $O(n \log n)$ \\ \hline
   \end{tabular}
   }
  \end{center}
  \caption{Summary of results.}
  \label{tab:complexity}
\end{table}

%% file: preliminaries_arxiv.tex
\section{Preliminaries}\label{sec:preliminaries}

Let $\Sigma$ be the \emph{alphabet}.
An element of $\Sigma^*$ is called a \emph{string}.
The length of a string $T$ is denoted by $|T|$.
String $s \in \Sigma^{*}$ is said to be a \emph{subsequence} of string
$T \in \Sigma^{*}$ if $s$ can be obtained by removing zero or more
characters from $T$.

For a string $T$ and an integer $1 \leq i \leq |T|$,
$T[i]$ denotes the $i$-th character of $T$.
For two integers $1 \leq i \leq j \leq |T|$,
$T[i..j]$ denotes the substring of $T$
that begins at position $i$ and ends at position $j$.
For convenience, let $T[i..j] = \varepsilon$ when $i > j$.

\subsection{k-(Sub-)Cadences}

The term ``cadence'' has been used in slightly different ways in the literature (e.g., see~\cite{Cadences_Gardelle, Cadences_Lothaire, Cadences_Amir}).
In this paper, we use the definitions of cadences and sub-cadences which are used in~\cite{Cadences_Amir} and~\cite{Funakoshi_STACS2020}.

For integers $i$ and $d$, the pair $(i,d)$ is called a
\emph{$k$-sub-cadence} of $T \in \Sigma^{n}$
if $T[i] = T[i+d] = T[i+2d] = \cdots = T[i+(k-1)d]$,
where  $1 \leq i \leq n$ and $1 \leq d \leq \lfloor \frac{n-1}{k-1}\rfloor$.
The set of $k$-sub-cadences of $T$ can be
defined as follows:

\begin{definition}
For $T\in \Sigma^{n}$, $n \in \mathcal{N}$, and
$k \in [1..n]$,
\begin{eqnarray*}
\ksc(T,k) = \left\{(i,d) \left|
\begin{array}{l}
T[i] = T[i+d] = T[i+2d] = \cdots = T[i+(k-1)d]\\
1 \leq i \leq n, 1 \leq d \leq \lfloor \frac{n-1}{k-1}\rfloor
\end{array}
\right.
\right\}.
\end{eqnarray*}
\end{definition}

For integers $i$ and $d$, the pair $(i,d)$ is called a
\emph{$k$-cadence} of $T \in \Sigma^{n}$
if $(i,d)$ is a $k$-sub-cadence and
satisfies the inequalities $i-d \leq 0$ and $n < i+kd$.
The set of $k$-cadences of $T$ can be
defined as follows:

\begin{definition}
For $T\in \Sigma^{n}$, $n \in \mathcal{N}$, and
$k \in [1..n]$,
\begin{eqnarray*}
\kc(T,k) = \left\{(i,d) \left|
\begin{array}{l}
T[i] = T[i+d] = T[i+2d] = \cdots = T[i+(k-1)d]\\
1 \leq i \leq n, 1 \leq d \leq \lfloor \frac{n-1}{k-1}\rfloor,
i-d \leq 0, n < i+kd
\end{array}
\right.
\right\}.
\end{eqnarray*}
\end{definition}

\subsection{Equidistant Subsequence Occurrences}
For integers $i$ and $d$, we say that pair $(i,d)$ is an
\emph{equidistant subsequence occurrence} of $P\in\Sigma^m$ in $T\in\Sigma^{n}$
if $P = T[i] \cdot T[i+d] \cdot T[i+2d] \cdots T[i+(m-1)d]$,
where  $1\leq i\leq n$ and $1 \leq d \leq \lfloor \frac{n-1}{m-1}\rfloor$.
The set of equidistant subsequence occurrences of $P$ in $T$ can be
defined as follows:

\begin{definition}
For $T\in \Sigma^{n}, P \in \Sigma^{m}$ and $n,m\in \mathcal{N}$,
\begin{eqnarray*}
\esp(T,P) = \left\{(i,d) \left|
\begin{array}{l}
P=T[i] \cdot T[i+d] \cdot T[i+2d] \cdots T[i+(m-1)d]\\
1\leq i \leq n, 1\leq d \leq \lfloor \frac{n-1}{m-1}\rfloor
\end{array}
\right.
\right\}.
\end{eqnarray*}
\end{definition}

\subsection{Equidistant Abelian Subsequence Occurrences}
Two strings $S_1$ and $S_2$ are said to be \emph{Abelian equivalent}
if $S_1$ is a permutation of $S_2$, or vice versa.
Now for integers $i$ and $d$, we say that pair $(i,d)$ is an
\emph{equidistant Abelian subsequence occurrence} of $P\in\Sigma^m$ in $T\in\Sigma^{n}$
if $T[i] \cdot T[i+d] \cdot T[i+2d] \cdots T[i+(m-1)d]$ and $P$ are Abelian equivalent,
where  $1\leq i\leq n$ and $1 \leq d \leq \lfloor \frac{n-1}{m-1}\rfloor$.
The set of equidistant Abelian subsequence occurrences of $P$ in $T$ can be
defined as follows:

\begin{definition}
For $T\in \Sigma^{n}, P \in \Sigma^{m}$ and $n,m\in \mathcal{N}$,
\begin{eqnarray*}
\easp(T,P) = \left\{(i,d) \left|
\begin{array}{l}
T[i] \cdot T[i+d] \cdots T[i+(m-1)d] {\ \rm and \ } P {\ \rm are \ Abelian \ equivalent}\\
1\leq i \leq n, 1\leq d \leq \lfloor \frac{n-1}{m-1}\rfloor
\end{array}
\right.
\right\}.
\end{eqnarray*}
\end{definition}

When it is clear from the context, we denote $\ksc(T,k)$ as $\ksc$, $\kc(T,k)$ as $\kc$, and $\esp(T,P)$ as $\esp$.

%% file: k-sub-cadences_arxiv.tex
\section{Detecting k-Sub-Cadences}\label{sec:k-sub-cadences}

In this section,
we consider algorithms for detecting $k$-sub-cadences.

\subsection*{Algorithm 1}
One of the most simple methods is as follows:
For each distance $d$ $(1 \leq d \leq \lfloor \frac{n-1}{k-1}\rfloor)$,
we construct text
$ST_d=T[1]\cdot T[1+d]\cdots T[1+d \lfloor \frac{n-1}{d}\rfloor]\cdot\$ \cdot T[2]\cdot T[2+d]\cdots T[2+d \lfloor \frac{n-2}{d}\rfloor]\cdot \$ \cdots T[d]\cdot T[2d]\cdots T[d \lfloor \frac{n}{d}\rfloor]$
of length $n+d-1$.
If we would like to find $k$-sub-cadences with $d$ in text $T$, we find concatenations of the same character of length $k$ as substrings in $ST_d$.
\begin{figure}[ht]
  \centering{
    \includegraphics[width=0.70\linewidth]{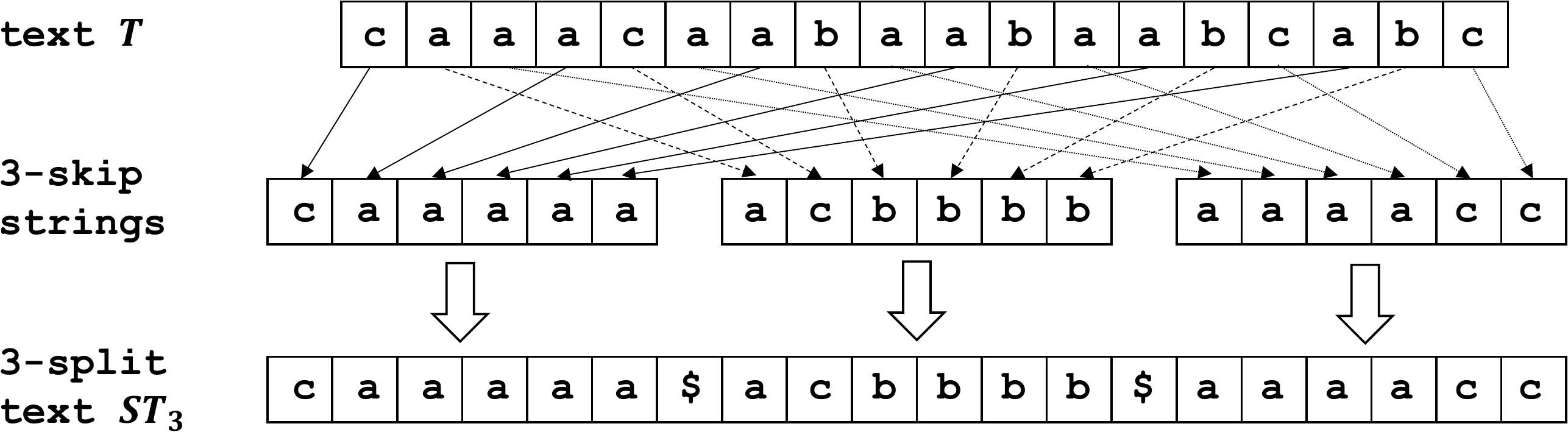}
  }
  \caption{Preprocessing for Algorithm 1.}
  \label{fig:3split}
\end{figure}

Fig.~\ref{fig:3split} is an example of 3-split text.
In this figure, the strings in the middle row are called \emph{$d$-skip strings}, and
the string on the bottom is called the \emph{$d$-split text $ST_d$}.
In $ST_d$, we use a symbol $\$\notin\Sigma$
in order to prevent detecting false occurrences concatenation of same character of the length $k$ across
the ends of $d$-skip strings as a $k$-sub-cadence.
The text obtained by concatenating all $ST_d$ for all $1\leq d \leq \lfloor
\frac{n-1}{k-1}\rfloor$ and $\$$ is called the \emph{split text}.
If we prepare the split text,
we can compute $\ksc$ simply by checking
that the same character is repeated $k$ times.

The length of $ST_d$ is at most $n+d$ including $\$$.
The maximum value of $d$ is $\lfloor \frac{n-1}{k-1}\rfloor$, and therefore, the
number of $ST_d$ of text $T$ is at most $\lfloor \frac{n-1}{k-1} \rfloor$.
Hence, the length of the split text of $T$ is $O(\frac{n^2}{k})$.
We can check that the same character is repeated $k$ times in the split text in $O(\frac{n^2}{k})$ time.
Although we have presented the split text to ease the description, it does not have to be constructed explicitly.

From the above, we can get the following result.
\begin{theorem} \label{theo:k-sub-1}
  There is an algorithm for locating all $k$-sub-cadences
  for given $k$ $(1 \leq k \leq n)$
  which uses $O\left(\frac{n^2}{k}\right)$ time and $O(n)$ space.
\end{theorem}

As can be seen from the example of $T=\mathtt{a}^n$, $|\ksc|$ can be $\Omega(\frac{n^2}{k})$.
Therefore, when we locate all $(i,d) \in \ksc$,
this algorithm is optimal in the worst case.
In the next subsection,
we show a counting algorithm that is efficient when the value of $k$ is small.
Moreover, we show a locating algorithm that is efficient
when both the value of $k$ and $|\ksc|$ is small.

\subsection*{Algorithm 2}

In this subsection, we will show the following result:
\begin{theorem} \label{theo:k-sub-2}
  For a constant size alphabet,
  there is an $O\left(\frac{n^2}{\log n}\right)$ time algorithm for counting all $k$-sub-cadences for given $k$.
  We can also locate these occurrences
  in $O\left(\frac{n^2}{\log n}+\occ\right)$ time,
  where $\occ$ is the number of the outputs.
  For a general ordered alphabet,
  there is an $O\left(\frac{n^2 \sqrt{k}}{\sqrt{\log n}}\right)$ time algorithm for counting all $k$-sub-cadences for given $k$.
  We can also locate these occurrences
  in $O\left(\frac{n^2 \sqrt{k}}{\sqrt{\log n}}+\occ\right)$ time.
  These algorithms run in $O(n)$ space.
\end{theorem}

Note that for counting all $k$-sub-cadences,
for a constant size alphabet (resp. for a general ordered alphabet),
this algorithm is faster than Algorithm 1
if $k$ is $o(\log n)$ (resp. $o\left(\sqrt[3]{\log n}\right)$).
For locating all $k$-sub-cadences,
for a constant size alphabet (resp. for a general ordered alphabet),
if $|\ksc|$ is $o(\frac{n^2}{k})$
and $k$ is $o(\log n)$ (resp. $o\left(\sqrt[3]{\log n}\right)$),
then this algorithm is faster.

Now we will show how to count all $k$-sub-cadences
of character $c \in\Sigma$.
Let $\delta_c[1..n]$ be a binary sequence for character $c$ defined as follows:
\[\delta_c[i] :=
\begin{cases}
   $1$ &\textup{if $T[i] = c$}, \\ $0$ &\textup{if $T[i] \neq c$}.
\end{cases}\]

\begin{figure}[ht]
  \centering{
    \includegraphics[width=0.70\linewidth]{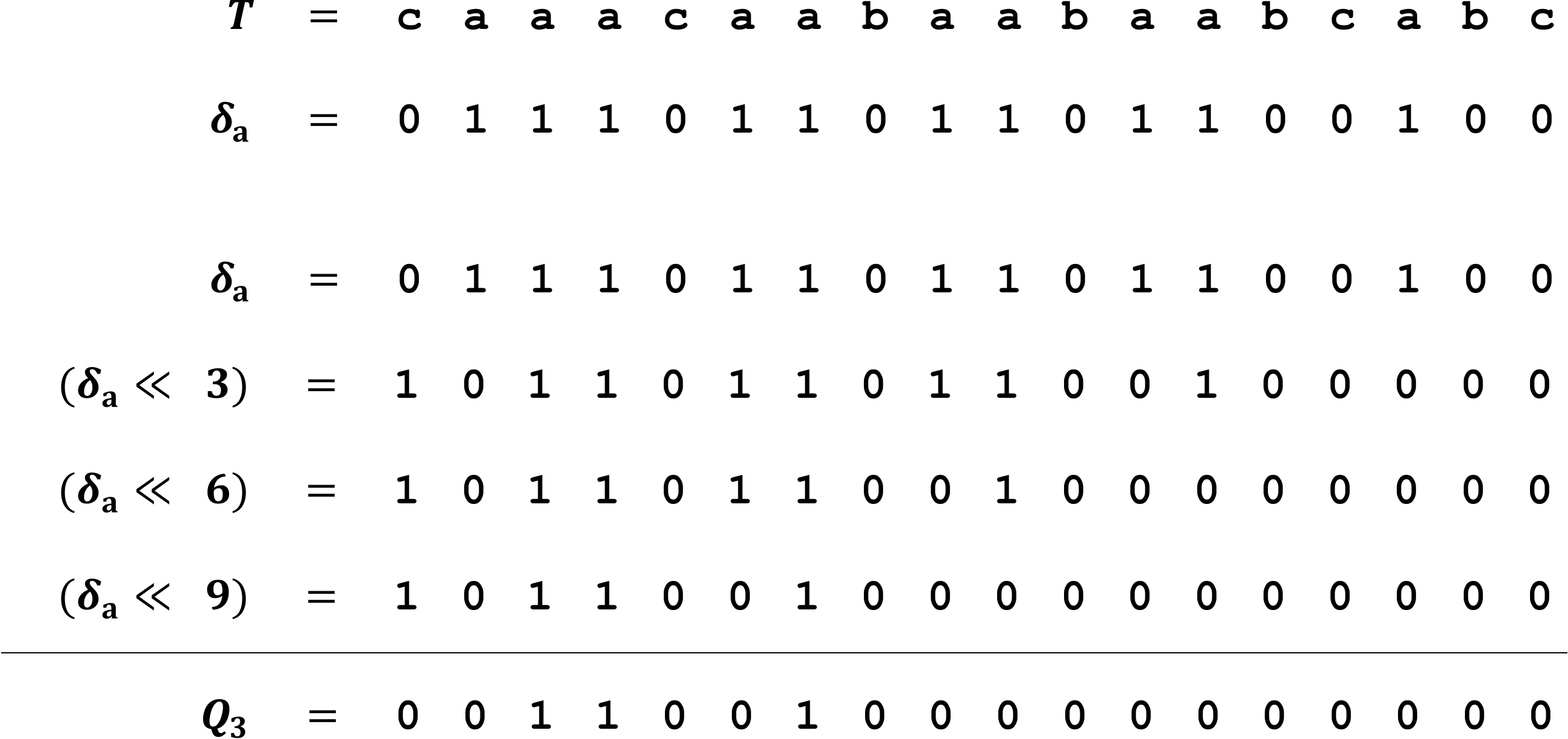}
  }
  \caption{Let $T=\mathtt{caaacaabaabaabcabc}$.
  $(3,3)$, $(4,3)$, and $(7,3)$ are $4$-sub-cadences of character `$\mathtt{a}$' with $d=3$.}
  \label{fig:alg2}
\end{figure}

If $(i,d)$ is a $k$-sub-cadence,
$\delta_c[i]=\delta_c[i+d]=\cdots=\delta_c[i+(k-1)d]=1$.
Therefore we can check whether $(i,d)$ is a $k$-sub-cadence or not by computing $\delta_c[i] \cdot \delta_c[i+d] \cdots \delta_c[i+(k-1)d]$.
To compute this, we use bit-parallelism, i.e, the bit-wise operations AND and SHIFT\_LEFT, denoted by
$\texttt{\&}$ and $\texttt{<<}$,
respectively, as in the C language.
For each $d$ $(1 \leq d \leq \lfloor \frac{n-1}{k-1} \rfloor)$,
let $Q_d = \delta_c \texttt{ \& } (\delta_c \texttt{ << } d) \texttt{ \& } (\delta_c \texttt{ << } 2d) \texttt{ \& } \cdots \texttt{ \& } (\delta_c \texttt{ << } (k-1)d)$.
If $Q_d[i]=1$, $(i,d)$ is a $k$-sub-cadence.
See Figure~\ref{fig:alg2} for a concrete example.

If we want to count all $k$-sub-cadences with $d$,
we only have to count the number of $1$'s in $Q_d$.
If we want to locate all $k$-sub-cadences with $d$,
we have to locate all $1$'s in $Q_d$.

In the word RAM model,
SHIFT\_LEFT and AND operations can be done in
constant time per operation on bit sequences of length $O(\log n)$.
Since $\delta_c$ is a binary sequence of length $n$,
one SHIFT\_LEFT or AND operation can be done in $O(\frac{n}{\log n})$ time.
Therefore, $Q_d$ can be obtained in $O(k \frac{n}{\log n})$ time.
Since it is known that the number of $1$'s in a bit sequence of length $O(\log n)$ can be obtained in $O(1)$ time by using the ``popcnt'' operation,
the number of $1$'s in $Q_d$ can be counted in $O(\frac{n}{\log n})$ time.
Hence, for all $1 \leq d \leq \lfloor \frac{n-1}{k-1} \rfloor$, we can count all $k$-sub-cadences of character $c$ in
$O\left(k \frac{n}{\log n} \lfloor \frac{n-1}{k-1} \rfloor + \frac{n}{\log n} \lfloor \frac{n-1}{k-1} \rfloor\right) \subseteq O(\frac{n^2}{\log n})$ time.
Also it is known that the position of the rightmost $1$ (the least significant set bit) in a bit sequence of length $O(\log n)$ can be answered in constant time.
We split $Q_d$ into $O(\frac{n}{\log n})$ blocks of length $O(\log n)$.
For each block,
the least significant set bit can be found in $O(1)$ time if the block contains at least one $1$.
After finding the least significant set bit,
we mask this bit to $0$ and do the above operation again.
Bit mask operation can be done in $O(1)$ time.
Hence, we can answer all the positions of $1$'s in $Q_d$
in $O(\frac{n}{\log n} + \occ)$ time.
Therefore, we can locate all $k$-sub-cadences of character $c$ in $O(\frac{n^2}{\log n} + \occ)$ time.

We showed how to detect all $k$-sub-cadences
of character $c$,
so we can detect all $k$-sub-cadences by doing the above operations for each character in $\Sigma$.
For a constant size alphabet,
since we only do the above operations a constant number of times,
we can count all $k$-sub-cadences in $O(\frac{n^2}{\log n})$ time.
We can also locate these occurrences in $O(\frac{n^2}{\log n} + \occ)$ time.
However, for a general ordered alphabet,
we have to do the above operations $|\Sigma|$ times.

For a general ordered alphabet,
if the number of occurrences of the character is small,
we use another algorithm that generalizes Amir et al.'s algorithm~\cite{Cadences_Amir} for detecting $3$-cadences to $k$-sub-cadences:
Let $N_c$ be the set of positions which are occurrences of a character $c$.
If we pick two positions in $N_c$ and regard the smaller one as the starting position $i$ of $k$-sub-cadences and the larger one as the second position $i+d$ of a $k$-sub-cadence, then the distance $d$ is uniquely determined.
We can check whether the pair $(i,d)$ is a $k$-sub-cadence or not in $O(k)$ time.
Since the number of pairs is at most $|N_c|^2$,
we can count or locate $k$-sub-cadences of character $c$ in $O(k|N_c|^2)$ time.

Thus, for a general ordered alphabet,
all $k$-sub-cadences can be counted in\\
$O(\sum_{c \in \Sigma}^{} \min\{k|N_c|^2, \frac{n^2}{\log n}\})$ time.
Since $O(\sum_{c \in \Sigma}^{} \min\{k|N_c|^2, \frac{n^2}{\log n}\})$ is maximized when\\ $k|N_c|^2=\frac{n^2}{\log n}$,
$O(\sum_{c \in \Sigma}^{} \min\{k|N_c|^2, \frac{n^2}{\log n}\}) \subseteq O(\sum_{c \in \Sigma}^{} \frac{n^2}{\log n}) \subseteq O((\sum_{c \in \Sigma}^{} |N_c|) \frac{n \sqrt{k}}{\sqrt{\log n}}) \subseteq O(\frac{n^2 \sqrt{k}}{\sqrt{\log n}})$. Therefore we can count in $O(\frac{n^2 \sqrt{k}}{\sqrt{\log n}})$ time by using Algorithm 2 and the generalized algorithm described above algorithm.
Also, all $k$-sub-cadences can be located in
$O(\sum_{c \in \Sigma}^{} \min\{k|N_c|^2, \frac{n^2}{\log n}+\occ_c\})$ time
where $\occ_c$ is the number of $k$-sub-cadences of character $c$.
Since $O(\sum_{c \in \Sigma}^{} \min\{k|N_c|^2, \frac{n^2}{\log n}+\occ_c\}) \subseteq O(\sum_{c \in \Sigma}^{} \min\{k|N_c|^2, \frac{n^2}{\log n}\}+\occ) \subseteq O(\frac{n^2 \sqrt{k}}{\sqrt{\log n}}+ \occ)$,
we can locate in $O(\frac{n^2 \sqrt{k}}{\sqrt{\log n}}+ \occ)$ time.

From the above, we obtain the following result:
\begin{theorem} \label{theo:k-sub-cadences}
  For a constant size alphabet (resp. for a general ordered alphabet),
  all $k$-sub-cadences with given $k$ can be counted in
  $O\left(\min\left\{\frac{n^2}{k},\frac{n^2}{\log n}\right\}\right)$ time\\ (resp. $O\left(\min\left\{\frac{n^2}{k},\frac{n^2 \sqrt{k}}{\sqrt{\log n}}\right\}\right)$ time) and $O(n)$ space,
  and can be located in\\
  $O\left(\min\left\{\frac{n^2}{k},\frac{n^2}{\log n}+\occ\right\}\right)$ time (resp. $O\left(\min\left\{\frac{n^2}{k},\frac{n^2 \sqrt{k}}{\sqrt{\log n}}+\occ\right\}\right)$ time) and $O(n)$ space.
\end{theorem}

%% file: k-cadences_arxiv.tex
\section{Detecting k-Cadences}\label{sec:k-cadences}

In this section,
we consider algorithms for detecting $k$-cadences.

\subsection*{Algorithm 3}
We use same techniques of Algorithm 1 and then check whether each element belongs to $\kc$ or not.
After computing $\ksc$ by using Algorithm 1,
we have to check if each $(i,d) \in \ksc$ satisfies
the following formulas:
$i \leq d$ and $i+kd > n$.
Since $|\ksc| \in O(\frac{n^2}{k})$,
we can do this operation in $O(\frac{n^2}{k})$ time.
Therefore, we can obtain the following result:
\begin{theorem} \label{theo:k-1}
  There is an algorithm for locating all $k$-sub-cadences
  for given $k$
  which uses $O\left(\frac{n^2}{k}\right)$ time and $O(n)$ space.
\end{theorem}

\subsection*{Algorithm 4}

Now, we will show the following result:
\begin{theorem} \label{theo:k-cadences-2}
  For a constant size alphabet,
  there is an $O\left(\frac{1}{\log n}\left(\frac{n^2}{k^2}+kn\right)\right)$ time algorithm for counting all $k$-cadences for given $k$.
  We can also locate these occurrences
  in\\ $O\left(\frac{1}{\log n}\left(\frac{n^2}{k^2}+kn\right)+\occ\right)$ time.
  For a general ordered alphabet,
  there is an \\$O\left(\frac{n \sqrt{k}}{\sqrt{\log n}}\sqrt{\frac{n^2}{k^2}+kn}\right)$ time algorithm for counting all $k$-sub-cadences for given $k$.
  We can also locate these occurrences
  in $O\left(\frac{n \sqrt{k}}{\sqrt{\log n}}\sqrt{\frac{n^2}{k^2}+kn}+\occ\right)$ time.
  These algorithms run in $O(n)$ space.
\end{theorem}

These time complexities are at least as fast as Algorithm 2 for $k$-sub-cadences.
Moreover, if the value of $k$ is neither constant nor $\Omega(n)$,
this algorithm is faster than Algorithm 2
because $\frac{n^2}{k^2}+kn$ will be $o(n^2)$.
Note that when we count all $k$-sub-cadences,
for a constant size alphabet,
this algorithm is faster than Algorithm 3
if $k$ is $o(\sqrt{n \log n})$.
(This is because $\frac{n}{k}+k^2$ will be $o(n \log n)$.)
Also, for a general ordered alphabet,
this algorithm is faster
if $k$ is $o(\log n)$.
(This is because $k \sqrt{k} \sqrt{\frac{n^2}{k^2}+kn}$ will be $o\left(n \sqrt{\log n}\right)$
and then $kn^2+k^4n$ will be $o(n^2 \log n)$.)
When we locate all $k$-sub-cadences,
for a constant size alphabet (resp. for a general ordered alphabet),
if $|\kc|$ is $o(\frac{n^2}{k})$
and $k$ is $o(\sqrt{n \log n})$ (resp. $o(\log n)$)
then this algorithm is faster.

First, we will show that (the size of) $\kc$ can be obtained by using the similar techniques of Algorithm 2 of a character, and then we will show how to speed up.

Again, each $(i,d)$ has to satisfy the following formulas:
$i \leq d$ and $i+kd > n$,
that is $n-kd < i \leq d$.
Then let $R_d[1..n]$ be the binary sequence defined as follows:
\[R_d[i] :=
\begin{cases}
   $1$ &\textup{if $n-kd < i \leq d$}, \\
   $0$ &\textup{otherwise}.
\end{cases}\]
Let $Q'_d = Q_d \texttt{ \& } R_d$.
If $Q'_d[i]=1$, $(i,d)$ is a $k$-cadence.
Since $R_d$ and $Q'_d$ can be computed in $O(\frac{n}{\log n})$ time,
this algorithm runs in the same time complexity as Algorithm 2.

Now we show how to speed up this algorithm.
In the above algorithm,
we obtain $Q'_d$ by masking $Q_d$.
However, to calculate $k$-cadences, we need only the range $[n-kd..d]$ of the sequence, and it is useless to calculate other ranges.
Therefore, we compute $Q'_d$ by the following operations:
$Q'_d = \delta_c[n-kd+1..d] \texttt{ \& } (\delta_c \texttt{ << } d)[n-kd+1..d] \texttt{ \& } (\delta_c \texttt{ << } 2d)[n-kd+1..d] \texttt{ \& } \cdots \texttt{ \& } (\delta_c \texttt{ << } (k-1)d)[n-kd+1..d] = \delta_c[n-kd+1..d] \texttt{ \& } \delta_c[n-(k-1)d+1..2d] \texttt{ \& } \delta_c[n-(k-2)d+1..3d] \texttt{ \& } \cdots \texttt{ \& } \delta_c[n-(k-m+1)d+1..md]$.
The length of range $[n-kd+1..d]$ is at most $(k+1)d-n$ with $\lceil \frac{n}{k+1} \rceil \leq d \leq \lfloor \frac{n-1}{k-1} \rfloor$.
We can compute all $Q'_d$ for $\lceil \frac{n}{k+1} \rceil \leq d \leq \lfloor \frac{n-1}{k-1} \rfloor$\\
in $\sum_{d=\lceil \frac{n}{k+1} \rceil}^{\lfloor \frac{n-1}{k-1} \rfloor}\left(\frac{(k+1)d-n}{\log n}k\right)$ time.
Then,
\footnotesize
\begin{align*}
  & \sum_{d=\left\lceil \frac{n}{k+1} \right\rceil}^{\left\lfloor \frac{n-1}{k-1} \right\rfloor}\left(\frac{(k+1)d-n}{\log n}k\right) \\
  & =\sum_{d=1}^{\left\lfloor \frac{n-1}{k-1} \right\rfloor}\left(\frac{(k+1)d-n}{\log n}k\right)-\sum_{d=1}^{\left\lceil \frac{n}{k+1} \right\rceil-1}\left(\frac{(k+1)d-n}{\log n}k\right) \\
  & =\frac{k}{\log n} \left( (k+1)\frac{\left\lfloor\frac{n-1}{k-1}\right\rfloor\left(\left\lfloor\frac{n-1}{k-1}\right\rfloor+1\right)}{2}-\left\lfloor\frac{n-1}{k-1}\right\rfloor n \right. \\ & \left. \qquad -(k+1)\frac{\left\lceil\frac{n}{k+1}\right\rceil\left(\left\lceil\frac{n}{k+1}\right\rceil-1\right)}{2} +\left(\left\lceil\frac{n}{k+1}\right\rceil-1\right)n\right) \\
  & < \frac{1}{2 \log n}\left( \frac{4n^2+n(k^3+k^2-5k-5)-k^3+3k+2}{(k-1)^2}+ 3nk +k^2 +k\right).
\end{align*}
\normalsize

$\frac{1}{2 \log n}\left( \frac{4n^2+n(k^3+k^2-5k-5)-k^3+3k+2}{(k-1)^2}+ 3nk +k^2 +k\right)$ is at most $O(\frac{1}{\log n}( \frac{n^2}{k^2}+kn))$.\\
Therefore, we can count all $k$-cadences of a character in $O(\frac{1}{\log n}( \frac{n^2}{k^2}+kn))$ time.
For a locating algorithm and for a general ordered alphabet,
we can use same techniques of the above section.
Therefore we get Theorem~\ref{theo:k-cadences-2}.

From the above, we obtain the following result:
\begin{theorem} \label{theo:k-cadences}
  For a constant size alphabet (resp. for a general ordered alphabet),
  all $k$-cadences with given $k$ can be counted in
  $O\left(\min\left\{\frac{n^2}{k},\frac{1}{\log n}\left(\frac{n^2}{k^2}+kn\right)\right\}\right)$ time\\ (resp. $O\left(\min\left\{\frac{n^2}{k},\frac{n \sqrt{k}}{\sqrt{\log n}}\sqrt{\frac{n^2}{k^2}+kn}\right\}\right)$ time) and $O(n)$ space,
  and can be located in\\
  $O\left(\min\left\{\frac{n^2}{k},\frac{1}{\log n}\left(\frac{n^2}{k^2}+kn\right)+\occ\right\}\right)$ time\\(resp. $O\left(\min\left\{\frac{n^2}{k},\frac{n \sqrt{k}}{\sqrt{\log n}}\sqrt{\frac{n^2}{k^2}+kn}+\occ\right\}\right)$ time) and $O(n)$ space.
\end{theorem}

%% file: equidistant_subsequence_pattern_matching_arxiv.tex
\section{Detecting Equidistant Subsequence Pattern}\label{sec:equidistant_subsequence_pattern}

In this section,
we consider algorithms for detecting equidistant subsequence pattern.

\subsection*{Algorithm 5}
We use similar techniques of Algorithm 1.
For each distance $d (1 \leq d \leq \lfloor \frac{n-1}{k-1}\rfloor)$,
we construct text $ST_d$.
After preparing the split text,
we can compute $\esp$ using existing substring pattern matching algorithms.
Since Knuth-Morris-Pratt
algorithm~\cite{KnuthMP77} runs in $O(n)$ time for a text of length $n$,
we obtain the following result:
\begin{theorem} \label{theo:ESPM-1}
  There is an algorithm for locating all equidistant subsequence occurrences
  for given pattern $P$ of length $m$
  which uses $O\left(\frac{n^2}{m}\right)$ time and $O(n)$ space.
\end{theorem}

Like $\ksc$, for text $T=\mathtt{a}^n$ and pattern $P=\mathtt{a}^m$,
$|\esp|$ can be $\Omega(\frac{n^2}{m})$.
Therefore, when we locate all $(i,d) \in \esp$,
this algorithm is optimal in the worst case.
In the next subsection,
we show a counting algorithm that is efficient when the value of $m$ is small.
And we show a locating algorithm that is efficient
when the value of $m$ and $|\esp|$ is small.

\subsection*{Algorithm 6}

Now we will show the following results:
\begin{theorem} \label{theo:ESPM-2}
  There is an algorithm for counting all equidistant subsequence occurrences
  which uses $O\left(\frac{n^2}{\log n}\right)$ time and $O\left(\frac{|\Sigma_P| n}{\log n}\right)$ space,
  where $\Sigma_P$ is the set of distinct characters in the given pattern $P$.
  We can also locate these occurrences
  in $O\left(\frac{n^2}{\log n}+\occ\right)$ time and $O\left(\frac{|\Sigma_P| n}{\log n}\right)$ space.
\end{theorem}

\begin{figure}[ht]
  \centering{
    \includegraphics[width=0.70\linewidth]{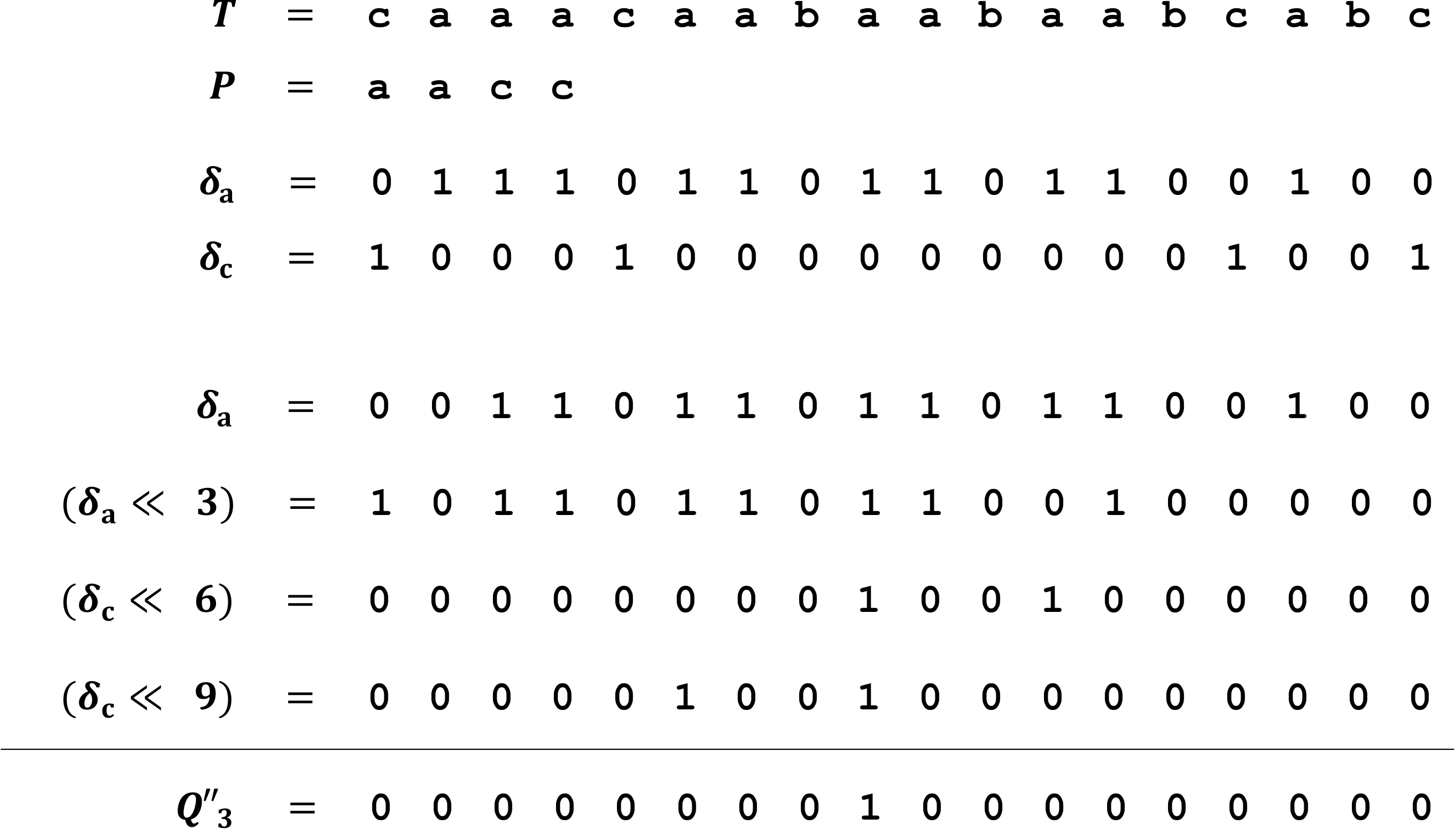}
  }
  \caption{Let $T=\mathtt{caaacaabaabaabcabc}$ and $P=\mathtt{aacc}$.
  $(9,3)$ is an occurrence of equidistant subsequence pattern with $d=3$.}
  \label{fig:alg6}
\end{figure}

First,
we construct $\delta_c$ for all $c \in \Sigma_P$.
For each $d (1 \leq d \leq \lfloor \frac{n-1}{m-1} \rfloor)$,
let $Q^{''}_d = \delta_{P[1]} \texttt{ \& }\\ (\delta_{P[2]} \texttt{ << } d) \texttt{ \& } (\delta_{P[3]} \texttt{ << } 2d) \texttt{ \& } \cdots \texttt{ \& } (\delta_{P[m]} \texttt{ << } (m-1)d)$.
If $Q^{''}_d[i]=1$, $(i,d)$ is a occurrence of equidistant subsequence pattern $P$.
See Figure~\ref{fig:alg6} for a concrete example.

All of the elements of $\esp$ can be counted / located by using a method similar to Algorithm 2 for $Q^{''}_d$.
After constructing $\delta_c$ for all $c \in \Sigma$,
all occurrences of equidistant subsequence pattern
can be counted in $O(\frac{n^2}{\log n})$ time and $O(n)$ space
and can be located in $O(\frac{n^2}{\log n}+\occ)$ time and $O(n)$ space.
Constructing $\delta_c$ for all $c \in \Sigma_P$
needs $O(\frac{|\Sigma_P| n}{\log n})$ time and space.
Since $\frac{|\Sigma_P| n}{\log n}$ is at most
$O(\frac{n^2}{\log n})$,
we get Theorem~\ref{theo:ESPM-2}.

If $m$ is $o(\log n)$,
Algorithm 6 is faster than Algorithm 5 and $O(\frac{|\Sigma_P| n}{\log n}) \subseteq O(n)$.
From the above, we obtain the following result:
\begin{theorem} \label{theo:ESPM}
  All occurrences of equidistant subsequence pattern
  can be counted in\\
  $O\left(\min\left\{\frac{n^2}{m},\frac{n^2}{\log n}\right\}\right)$ time and
  $O(n)$ space
  and can be located in
  $O\left(\min\left\{\frac{n^2}{m},\frac{n^2}{\log n}+\occ\right\}\right)$ time and $O(n)$ space.
\end{theorem}

%% file: equidistant_subsequence_pattern_of_length_three_arxiv.tex
\section{Detecting Equidistant Subsequence Pattern of Length Three}\label{sec:equidistant_subsequence_pattern_of_length_three}

In this section,
we show more efficient algorithms that count all occurrences of an equidistant subsequence pattern
for the case where the length of the pattern is three.
In addition,
we show an algorithm for counting all occurrences of equidistant Abelian subsequence patterns of length three.
Since we heavily use the techniques of~\cite{Funakoshi_STACS2020} for $3$-sub-cadences,
we first show their algorithm for $3$-sub-cadences and then generalize it for solving the equidistant subsequence pattern matching problem.

\subsection*{Counting 3-sub-cadences~\cite{Funakoshi_STACS2020}}
Let $a[1..n]$ and $b[1..n]$ be two sequences.
The sequence $c[1..2n]$ can be computed by the discrete acyclic convolution
$c[z] = \sum_{\substack{x+y=z \\ (x,y) \in [0,1,2,\dots,n]^2}} a[x] b[y]$.
The discrete acyclic convolution can be computed in
$O(n \log n)$ time by using the fast Fourier transform.
This convolution can be interpreted geometrically as follows:
$c[z] = \sum_{\substack{x+y=z \\ (x,y) \in G \cap \mathbb{Z}^2}} a[x] b[y]$,
where $G$ is the square given by $\{(x,y): 0\leq x,y \leq n \}$.

Funakoshi and Pape-Lange~\cite{Funakoshi_STACS2020} showed that
$3$-sub-cadences can be counted by using the discrete acyclic convolution.
If $(i,d)$ is a $3$-sub-cadence with a character $c$,
$\delta_c[i] \cdot \delta_c[i+2d]=1$ and $T[i+d]=c$.
Let $c[2z]=\sum_{\substack{x+y=2z \\ (x,y) \in [0,1,2,\dots,n]^2}} \delta_c[x] \delta_c[y]$,
then $c[2z]$ will be the number of the index $z$ lies in the middle of two $c$.
Since $z+z=2z$ and $\delta_c[z] \cdot \delta_c[z]=1$
if $T[z]=1$,
$c[2z]$ counts one false positive.
In addition,
$x+y=z$ and $\delta_c[x] \cdot \delta_c[y]=1$
if $x \neq y$,
$c[2z]$ counts twice for same $x$ and $y$.
Let $f[z]$ be the number of all $3$-sub-cadences
with a character $c$ such that the second occurrences of $c$ has index $z$.
$f[z]$ can be computed in $O(n \log n)$ time as follows:
\[f[z] := \begin{cases}
\frac{c[2z]-1}{2} &\textup{if $T[z]=c$},\\
0 &\textup{if $T[z]\neq c$}.
\end{cases}\]

Furthermore, they extended the geometric interpretation of convolution and showed that
if $G$ is a triangle with perimeter $p$,
the sequence $c$
can be computed in $O(p \log^2 p)$.

\subsection*{Counting Equidistant Subsequence Patterns of Length Three}

Now we show the algorithm for counting all occurrences of equidistant subsequence pattern whose length is three.
Let $g[z]$ be the number of all occurrences of
equidistant subsequence pattern
such that the second occurrences of $P$ has index $z$.

If $P=\alpha \alpha \alpha$,
this problem is equal to the counting all $3$-sub-cadences problem.
Therefore,
$g[z]$ can be computed in $O(n \log n)$ time as follows:
\[g[z] := \begin{cases}
\frac{c[2z]-1}{2} &\textup{if $T[z]=\alpha$}\\
0 &\textup{if $T[z]\neq \alpha$}
\end{cases}\]
where $c[2z]=\sum_{\substack{x+y=2z \\ (x,y) \in [0,1,2,\dots,n]^2}} \delta_\alpha[x] \delta_\alpha[y]$.

If $P=\alpha \beta \alpha$,
since the pattern is symmetrical,
$g[z]$ can be computed in $O(n \log n)$ time as follows,
by using almost the same technique as above:
\[g[z] := \begin{cases}
\frac{c[2z]}{2} &\textup{if $T[z]=\beta$}\\
0 &\textup{if $T[z]\neq \beta$}
\end{cases}\]
where $c[2z]=\sum_{\substack{x+y=2z \\ (x,y) \in [0,1,2,\dots,n]^2}} \delta_\alpha[x] \delta_\alpha[y]$.

However, if $P=\alpha \beta \gamma$,
$c[2z]=\sum_{\substack{x+y=2z \\ (x,y) \in [0,1,2,\dots,n]^2}} \delta_\alpha[x] \delta_\gamma[y]$
would also include
occurrences of equidistant subsequence pattern
$\gamma \beta \alpha$.
Thus, in order to compute $g[z]$,
we further add the condition $x<y$.
By using triangle convolution of~\cite{Funakoshi_STACS2020},
$g[z]$ can be computed in $O(n \log^2 n)$ time
as follows:
\[g[z] := \begin{cases}
c[2z] &\textup{if $T[z]=\beta$}\\
0 &\textup{if $T[z]\neq \beta$}
\end{cases}\]
where $c[z] = \sum_{\substack{x+y=z \\ (x,y) \in G \cap \mathbb{Z}^2}} \delta_\alpha[x] \delta_\gamma[y]$, where $G$ is the triangle as following figure~\ref{fig:esp3}.
\begin{figure}[ht]
  \centering{
    \includegraphics[width=0.235\linewidth]{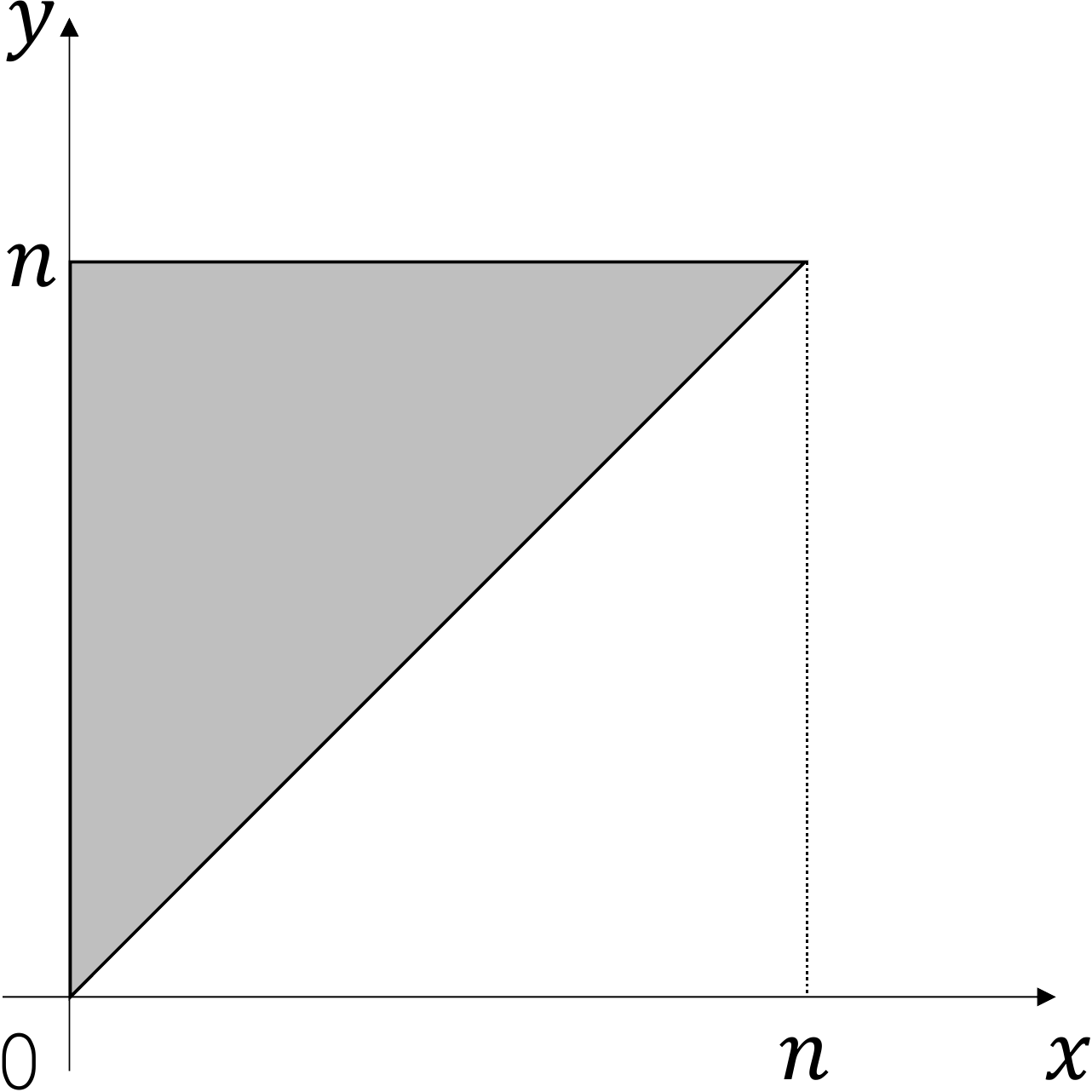}
  }
  \caption{The triangle $G$.}
  \label{fig:esp3}
\end{figure}

If $P=\alpha \alpha \gamma$ or $P=\alpha \gamma \gamma$,
we can compute $g[z]$ by using the same technique
as for the case of $P=\alpha \beta \gamma$.
Therefore, we get the following result:
\begin{theorem} \label{theo:esp3}
  All occurrences of equidistant subsequence pattern
  of length three
  can be counted in
  $O(n \log^2 n)$ time and $O(n)$ space.
\end{theorem}

\subsection*{Counting Equidistant Abelian Subsequence Patterns of Length Three}

Now we show the algorithm for counting all occurrences of equidistant Abelian subsequence pattern whose length is three.
In this subsection we consider the case where all of the three characters are distinct, namely, $P=\alpha \beta \gamma$.
The other cases can be computed similarly.

In the previous subsection,
we showed that
if $P=\alpha \beta \gamma$,
$c[2z]=\sum_{\substack{x+y=z \\ (x,y) \in [0,1,2,\dots,n]^2}} \delta_\alpha[x] \delta_\gamma[y]$ counts
the occurrences of equidistant subsequence pattern
$\gamma \beta \alpha$.
Therefore, we can compute all occurrences of equidistant subsequence pattern
$\alpha \beta \gamma$, $\gamma \beta \alpha$,
$\beta \gamma \alpha$, $\alpha \gamma \beta$,
$\gamma \alpha \beta$, and $\beta \alpha \gamma$
by using discrete acyclic convolution
for $P=\alpha \beta \gamma$, $P=\beta \gamma \alpha$, and $P=\gamma \alpha \beta$.
Hence, we can get following result:
\begin{theorem} \label{theo:easp3}
  All occurrences of equidistant Abelian subsequence pattern
  of length three
  can be counted in
  $O(n \log n)$ time and $O(n)$ space.
\end{theorem}